\documentclass[12pt,dvips]{article}
\usepackage[dvips]{graphicx}
\usepackage{graphicx}

\graphicspath{{plots/}}
\DeclareGraphicsExtensions{.eps.gz,.eps,.ps,.ps.gz}

\parskip=\itemsep               
\newcommand{\lwig}{\mbox{\,\raisebox{.3ex}
    {$<$}$\!\!\!\!\!$\raisebox{-.9ex}{$\sim$}\,}}
\newcommand{\gwig}{\mbox{\,\raisebox{.3ex}
    {$>$}$\!\!\!\!\!$\raisebox{-.9ex}{$\sim$}}\,}
\newcommand{\lambdabar}{{\hbox{$\lambda_e$\kern-1.9ex\raise+0.45ex\hbox{--}
\kern+0.2ex}}}

\newif\ifhepph

\hepphtrue

\ifhepph\date{\empty}\fi

\title{
\ifhepph{\normalsize
\rightline{DESY 03-057}\rightline{hep-ph/0306106}
      }\fi
\vskip 1cm 
\bf Production and detection of very light bosons\\ in the HERA tunnel 
       \vspace{21mm}} 
\author{A. Ringwald\\[4mm] 
Deutsches Elektronen-Synchrotron DESY, Hamburg, Germany}

\begin{document}
\begin{titlepage} 
  \maketitle
\vspace{3cm}
\begin{abstract}
There are strong theoretical arguments in favour of the existence
of very light scalar or pseudoscalar particles beyond the Standard 
Model which have, so far, remained undetected, due to their very weak coupling 
to ordinary matter. We point out that after HERA has been 
decommissioned, there arises a unique opportunity for 
searches for such particles: a number of HERA's four hundred 
superconducting dipole magnets might be recycled and used 
for laboratory experiments to produce and detect light neutral
bosons that couple to two photons, such as the axion. 
We show that, in this way, laser experiments searching for
photon regeneration or polarization effects 
in strong magnetic fields can
reach a sensitivity which is unprecedented in pure laboratory 
experiments and exceeds astrophysical limits from stellar evolution considerations.
\end{abstract}


\thispagestyle{empty}
\end{titlepage}
\newpage \setcounter{page}{2}

There are strong theoretical arguments in favour of the existence
of very light scalar or pseudoscalar particles beyond the Standard 
Model which have, so far, 
remained undetected, due to their very weak coupling 
to ordinary matter. 
They arise if there is a global continuous symmetry in the theory 
that is spontaneously broken in the vacuum. 

A prominent example is the axion  ($A^0$)~\cite{Weinberg:1978ma
}, which 
arises from a natural solution to the strong $CP$ problem.
The axion appears as a pseudo Nambu-Goldstone boson of a spontaneously broken Peccei-Quinn
symmetry~\cite{Peccei:1977hh
}, whose scale $f_A$ determines the mass,
\begin{equation}
\label{eq:ax_mass}
{ m_A} = { 0.62\times 10^{-3} \  {\rm eV}}\  
         \left( 
         10^{10}\ {\rm GeV}/ f_A  
        \right)\,, 
\end{equation}
and suppresses the coupling to Standard Model particles, $\propto 1/f_A$. 
The original axion model, with $f_A\sim v = 247$ GeV being of the order of 
the scale of electroweak symmetry breaking, is experimentally excluded 
(see e.g. Ref.~\cite{Hagiwara:fs}
and references therein), however so-called invisible axion 
models~\cite{Kim:1979if,
Zhitnitsky:1980tq
}, 
where $f_A\gg v$, are still allowed. Moreover, the invisible axion with 
$f_A\sim 10^{12}$~GeV seems to be a good candidate for the
cold dark matter component of the universe~\cite{Preskill:1982cy
}.     

\begin{figure}[t]
\vspace{-3.2cm}
\begin{center}
\includegraphics*[width=16cm]{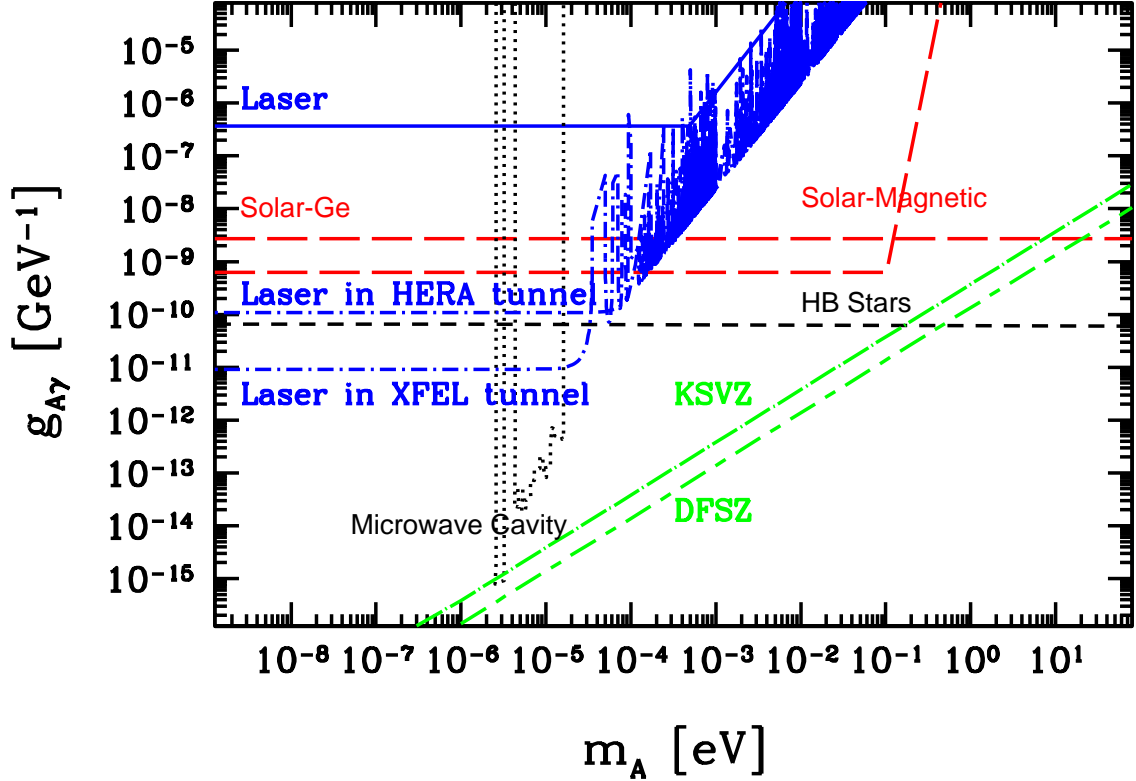}
\vspace{-2.5cm}
\caption[...]{Exclusion region in mass $m_A$ vs. axion-photon coupling $g_{A\gamma}$ 
for various 
current experiments (adapted from Ref.~\cite{Hagiwara:fs}, where also a detailed bibliography 
can be found) and for the ones proposed in this Letter (labelled as ``Laser in HERA tunnel'' and 
``Laser in XFEL tunnel'').
The laser experiments aim at both, 
axion production and detection in the laboratory~\cite{Semertzidis:1990qc,Ruoso:1992nx,Cameron:mr}.  
The microwave cavity experiments aim at axion detection under the assumption that axions are 
the galactic dark matter~\cite{DePanfilis:1987dk
}, the telescope search looks for axions thermally produced in 
galaxy clusters~\cite{Bershady:1990sw
}, and the solar-magnetic~\cite{Lazarus:1992ry
} and solar-Germanium~\cite{Avignone:an
} experiments search for axions from the sun. 
The constraint from helium burning (HB) stars arises from a consideration of the 
energy losses associated with axion production and the corresponding influence on stellar 
evolution~\cite{Raffelt:1985nk
}.
It is also shown that 
two quite distinct invisible axion models, namely the KSVZ~\cite{Kim:1979if
} 
(or hadronic) 
and the DFSZ~\cite{Zhitnitsky:1980tq
} (or grand unified) one, 
lead to quite similar axion-photon couplings.    
\hfill
\label{fig:ax_ph}}
\end{center}
\end{figure}

Clearly, it is of great interest to set stringent constraints on the properties
of such a light pseudoscalar. 
The interactions of axions and similar light pseudoscalars with Standard Model particles are model 
dependent, i.e. not a function of $1/f_A$ only.  
The most stringent constraints to date come from their coupling to photons, 
$g_{A\gamma}$, which arises via the axial anomaly, 
\begin{equation}
\label{eq:ax_ph}
        {\mathcal L}_{\rm int} = 
\frac{1}{4}\,{ g_{A\gamma}}\,A^0\ F_{\mu\nu} \tilde{F}^{\mu\nu} 
=
- { g_{A\gamma}}\,A^0\ {\mathbf E}\cdot {\mathbf B}\, ;
        \hspace{5ex}
        { g_{A\gamma}} = \frac{\alpha}{2\pi { f_A}} 
        \left( { \frac{E}{N}} -1.92\pm 0.08\right) \,,
\end{equation}
where $F_{\mu\nu}$ ($\tilde{F}^{\mu\nu}$) is the (dual) electromagnetic field strength tensor, 
$\alpha$ is the fine-structure constant, and 
$E/N$ is the ratio of electromagnetic over color anomalies, a model-dependent
ratio of order one. 
As illustrated in Fig.~\ref{fig:ax_ph}, which displays the 
axion-photon coupling~(\ref{eq:ax_ph}), in terms of its mass~(\ref{eq:ax_mass}), 
the strongest constraints on $g_{A\gamma}$ currently arise from cosmological and astrophysical 
considerations. They rely on axion production 
in cosmological or astrophysical environments.
Only the laser experiments quoted in Fig.~\ref{fig:ax_ph} aimed also at the production 
of axions in the laboratory\footnote{For other proposals to produce and detect very light axions in the
laboratory see e.g. Ref.~\cite{deRujula:1988mq
}.}.  

Let us discuss such laser experiments in some detail, since our proposal in this Letter 
rests on a scaled-up version of those.

\begin{figure}[t]
\begin{center}
\includegraphics*[bbllx=86pt,bblly=637pt,bburx=298pt,%
bbury=707pt,width=10cm,clip=]{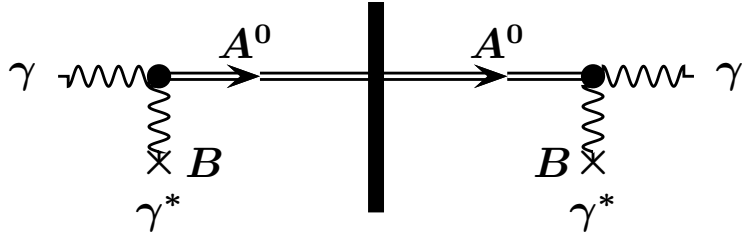}
\caption[...]{Schematic view of axion production through photon conversion 
in a magnetic field (left), subsequent travel through a wall, and final 
detection through photon regeneration (right). 
\hfill
\label{fig:light_wall}}
\end{center}
\end{figure}

The most straightforward and direct proposal is to set up a  photon regeneration experiment. 
It may be based on the 
idea~\cite{Anselm:1985gz,Gasperini:da,VanBibber:1987rq}\footnote{For variants of this idea, 
see Ref.~\cite{Hoogeveen:1990ep
}.} to send a polarized laser beam, with average power $\langle P\rangle$ and frequency $\omega$, 
along a superconducting dipole magnet of length $\ell$,  
such that the laser polarization is parallel to the magnetic field 
($\mathbf E_{\rm Laser}\, ||\, \mathbf B$). 
In this external magnetic field the photons may convert into axions via 
a Primakoff process (cf. Eq.~(\ref{eq:ax_ph}) and Fig.~\ref{fig:light_wall})
with a probability (see also Refs.~\cite{Sikivie:ip,
Raffelt:1987im}) 
\begin{equation}
P_{\gamma\rightarrow A} =
\frac{1}{4} \left( g_{A\gamma}\,{B}\,\ell\right)^2 F^2 (q\,\ell )\,, 
\end{equation}
where $q = m_A^2/(2\,\omega )\,(\ll m_A )$ is the momentum transfer to the magnet and   
\begin{equation}
F(q\,\ell ) = \frac{\sin\left(\frac{1}{2} q\,\ell \right)}{\frac{1}{2} q\,\ell }
\approx 1  
\end{equation}
is a form factor appropriate for the magnetic region of rectangular shape considered. Here, the 
approximate expression follows from expanding the sine which is
valid for light axions, 
$m_A^2\,\ell/(4\,\omega ) \ll \pi/2$.
If another identical dipole magnet is set up in line with the first magnet, with a sufficiently thick
wall between them to absorb completely the incident laser photons, 
then photons may be regenerated from the pure axion beam
in the second magnet (cf. Fig.~\ref{fig:light_wall}) with a probability 
$P_{A\rightarrow\gamma}=P_{\gamma\rightarrow A}$ and detected with
an efficiency $\epsilon$.  
The expected counting rate $R$ of such an experiment is given by~\cite{Ruoso:1992nx,Cameron:mr,VanBibber:1987rq}
\begin{equation}
\label{eq:ax_counting_rate}
R \equiv \frac{{\rm d}N_\gamma}{{\rm d}t} =  { \frac{\langle P\rangle }{\omega}}\ 
\frac{N_r}{2}\,P_{\gamma\rightarrow A}\ 
P_{A\rightarrow\gamma}\ \epsilon         
\,,
\end{equation}
if one makes use of the possibility of putting the first magnet into an 
optical cavity, where $N_r$ is the total number of reflections in the cavity.
For a sufficiently light axion, 
\begin{equation}
\label{eq:ax_coh}
{ m_A} 
\ll 
\sqrt{\frac{2\,\pi\,\omega}{\ell}}
=
{ 1.1\times 10^{-4}\ {\rm eV}}
       \ 
\,\left( \frac{{ \omega}}{1\ {\rm eV}} 
       \frac{100\ {\rm m}}{\ell} \right)^{1/2}
\,,
\end{equation}
the expected counting rate for a photon regeneration experiment according to 
an arrangement like in Fig.~\ref{fig:light_wall} is independent of the axion mass and 
a null experiment would thus yield a 90\,\% confidence level upper limit of\footnote{The same 
limit can be set on a possible coupling of a very light scalar particle ($\sigma^0$) 
to two photons, 
${\mathcal L}_{\rm int} = 
\frac{1}{4}\,{ g_{\sigma\gamma}}\,\sigma^0\ F_{\mu\nu} {F}^{\mu\nu} 
=
{ - g_{\sigma\gamma}}\,\sigma^0\ ({\mathbf E}^2 - {\mathbf B}^2)$, 
if one employs a laser polarization perpendicular to the magnetic field.}
\begin{equation}
\label{sensit}
g_{A\gamma} < 
\left( 1.8\times 10^{-10}\ {\rm GeV}^{-1}\right) 
\left( 
\frac{5\ {\rm T}}{{B}}\,
\frac{100\ {\rm m}}{\ell}\,
\right) 
\left( 
\frac{\omega}{1\ {\rm eV}}\,
\frac{10\ {\rm W}}{\langle P\rangle}\,
\frac{500}{N_r}
\frac{100\ {\rm d}}{t}\,
\frac{1}{\epsilon}
\right)^{1/4}
\,,
\end{equation}
in the mass range~(\ref{eq:ax_coh}). 
Indeed, a pilot photon regeneration experiment~\cite{Ruoso:1992nx,Cameron:mr}, employing for 
$t=220$~minutes an optical laser
of wavelength $\lambda =2\pi/\omega = 514$~nm and power $\langle P\rangle = 3$~W in an optical 
cavity with $N_r=200$, and using two superconducting 
dipole magnets with $B = 3.7$ T and $\ell = 4.4$ m, found, taking into account a 
detection efficiency of 
$\epsilon =0.055$, a $2\,\sigma$ upper limit of 
$g_{A\gamma}<6.7\times 10^{-7}$~GeV$^{-1}$ for axion-like pseudo-scalars with
mass $m_A<8\times 10^{-4}$ eV, in accordance with the expectation~(\ref{sensit}). 

Clearly, with considerable effort, a photon regeneration experiment can reach a sensitivity in 
$g_{A\gamma }$ which is comparable and even superior to the one obtained from stellar 
evolution, $g_{A\gamma}\,\lwig\, 6\times 10^{-11}$~GeV$^{-1}$ 
(constraint labelled ``HB Stars'' in Fig.~\ref{fig:ax_ph}). 
This has been emphasized in Eq.~(\ref{sensit}) by the choice of quite demanding benchmark 
parameters\footnote{The benchmark parameters for the laser ($\langle P\rangle$) and the cavity 
($N_r$) in Eq.~(\ref{sensit}) are similar to the ones used in gravitational wave 
detectors~\cite{Zawischa:ds
}.}, 
in particular for the linear extension $\ell$ of each of the regions before and behind the wall 
which are endowed with a magnetic field.

\begin{figure}
\begin{center}
\includegraphics*[bbllx=6pt,bblly=18pt,bburx=501pt,bbury=495pt,width=7.7cm]{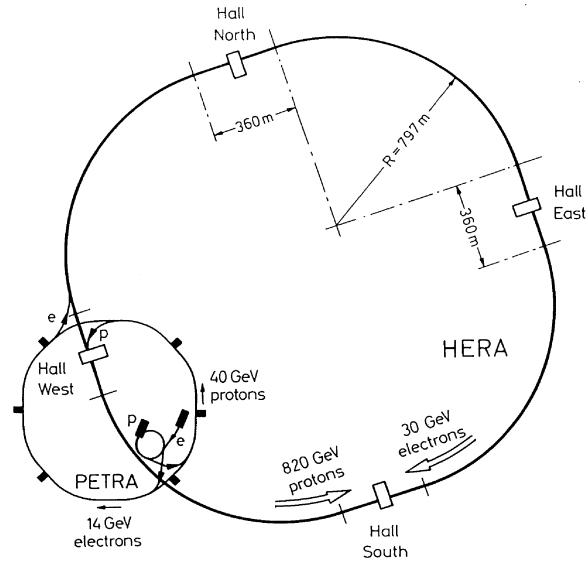}
\caption[...]{The Hadron Elektron Ring Anlage HERA at DESY in Hamburg.
The HERA tunnel has four straight sections, each of length $\approx 360$~m, at the  
location of the present experiments.  
\hfill
\label{fig:hera}}
\end{center}
\end{figure}
\begin{figure}
\begin{center}
\includegraphics*[angle=180,width=8.5cm,clip=]{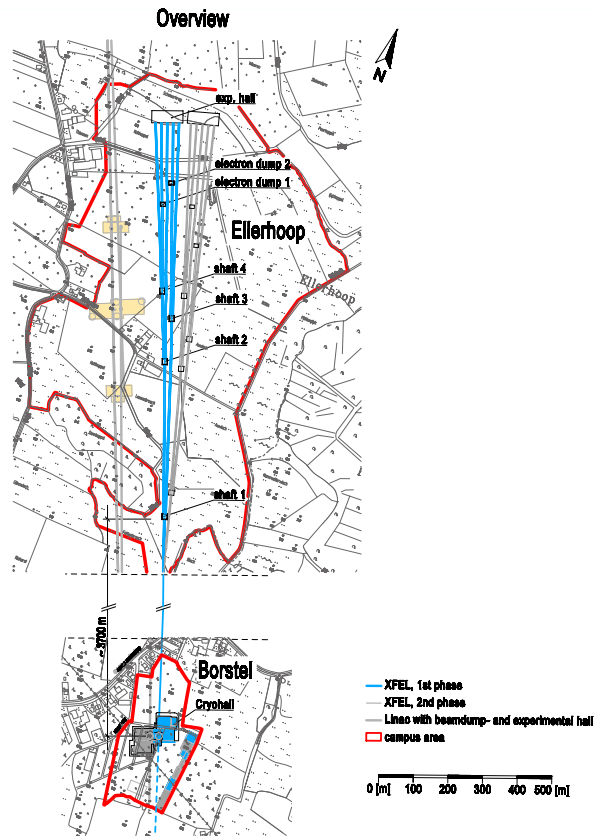}
\caption[...]{The TESLA XFEL campus North-West of the DESY laboratory~\cite{Materlik:2001qr}, whose
commissioning is expected in 2011. 
The XFEL electron beam is accelerated by a dedicated $20$~GeV superconducting linear accelerator 
starting at a supply hall $\approx 4$~km south of the XFEL laboratory. 
\hfill
\label{fig:xfel_camp}}
\end{center}
\end{figure}

Although it sounds, at first sight, quite unrealistic to achieve these benchmark parameters 
in the foreseeable future, there is, indeed, a quite realistic opportunity even to surpass 
them: in the not so far distant future, possibly already by the end of $2006$, the electron positron collider 
HERA at DESY in Hamburg (cf. Fig.~\ref{fig:hera}) will 
be decommissioned and the $\approx 400$ superconducting dipole magnets, each of which 
achieving $B=5$~T and having a length of $10$~m,  can, in principle, 
be recycled and used for a photon regeneration experiment.
One may envisage to employ $\approx 17+17$ of those magnets aligned along one of the 
four straight sections of HERA (cf. Fig.~\ref{fig:hera}), to supply a magnetic field of 
$5$~T over a linear extension of $2\,\ell\approx 340$~m. With such an arrangement one will
be able to reach a sensitivity of $g_{A\gamma}\,\lwig\,1\times 10^{-10}$~GeV$^{-1}$ 
for $m_A\,\lwig\,10^{-4}$~eV 
(labelled ``Laser in HERA tunnel'' in  Fig.~\ref{fig:ax_ph}), which is unprecedented in pure 
laboratory experiments,  
similar to the limit from stellar evolution, and competes with the designed sensitivity
of the CERN Axion Solar Telescope CAST~\cite{Zioutas:1998cc
}, $g_{A\gamma}\,\lwig\, (5\div 9)\times 10^{-11}$ GeV$^{-1}$ for $m_A\,\lwig\,10^{-2}$~eV, 
which looks for axions from the sun. 

In a later stage, one may think on deploying all $400$ decommissioned HERA 
dipole magnets ($2\,\ell\approx 4000$~m) in the $4$~km long TESLA XFEL tunnel at DESY, 
in which after $2010$ a superconducting linear accelerator will run to provide high-quality electron bunches
for the X-ray free electron lasers (XFELs) in a dedicated laboratory~\cite{Materlik:2001qr} 
(cf. Fig.~\ref{fig:xfel_camp}).  
The corresponding sensitivity, $g_{A\gamma}\,\lwig\,9\times 10^{-12}$~GeV$^{-1}$ 
for $m_A\,\lwig\,3\times 10^{-5}$~eV (labelled ``Laser in XFEL tunnel'' in  Fig.~\ref{fig:ax_ph}),  
has so far been only probed by microwave cavity searches for axions, under
the assumption that they are the dominant part of the galactic cold dark matter 
(cf. Fig.~\ref{fig:ax_ph}). 

Both proposals, the first stage photon regeneration 
experiment in the HERA tunnel and
the second stage in the TESLA XFEL tunnel, share the advantage that the 
necessary infrastructure for running superconducting dipole magnets is already
available at these sites. 
The sensitivity of both proposals, however, still does not extend to larger axion masses. 
Correspondingly, standard axion couplings (labelled ``KSVZ'' and ``DFSZ'' in Fig.~\ref{fig:ax_ph}) 
are beyond the reach of them. In order to probe larger axion masses, one may  
segment the $N=17\div 200$ aligned dipole magnets into $n$ subgroups of 
alternating polarity~\cite{VanBibber:1987rq} such that the appropriate form factor,
\begin{equation}
F(q\,\ell ) = \frac{\sin\left(\frac{1}{2} q\,\ell \right)}{\frac{1}{2} q\,\ell }
\,\tan (q\,\ell/(2\,n)) 
\,, 
\end{equation}
peaks at a nonzero value of $q$, thereby giving sensitivity to higher-mass (pseudo-)scalars.
With such an arrangement one may gain up to an order of magnitude in mass -- still
remaining, however, below standard axion values. 

Even larger axion masses may in principle be probed 
by the usage of the photon beam from one of the future X-ray free electron lasers
($\omega\approx 1$~keV in Eq.~(\ref{eq:ax_coh})) instead of the one from an optical laser  
($\omega\approx 1$~eV)~\cite{Ringwald:2001cp}. However, the poor longitudinal coherence
of the first generation XFEL beams limits the immediate prospects of their application 
to axion production in an external magnetic field. They will have a longitudinal coherence length 
$\ell_c = (0.5\div 30)$~$\mu$m~\cite{Materlik:2001qr} which is much 
less than the linear extension of the magnetic field $\ell = (170\div 2000)$~m. Under
such circumstances, the expected counting rate~(\ref{eq:ax_counting_rate}), which tacitly assumes 
$\ell_c>\ell$, diminishes by 
a factor $(\ell_c/\ell)^2$ and, correspondingly, the sensitivity~(\ref{sensit}) gets worse by a factor 
$(\ell/\ell_c)^{1/2}\approx 10^4$~\cite{VanBibber:1987rq}. 
For example, exploiting the photon beam from the 2-stage SASE 2 XFEL in the
TESLA XFEL laboratory~\cite{Materlik:2001qr}, with $\ell_c=30$~$\mu$m, $\omega = 14.4$~keV,   
$\langle P\rangle =100$~W, and $N_r=2$, one obtains, for $B=5$\,T and $\ell = 2$~km, 
a sensitivity of $g_{A\,\gamma}\,\lwig\,2\times 10^{-6}$~GeV$^{-1}$, for $m_A<3\times 10^{-3}$~eV. 
This is even worse than the current limit from the above mentioned pilot photon regeneration 
experiment (cf. Fig.~\ref{fig:ax_ph}), but extends of course to higher masses.  
Competitive limits from the usage of XFELs can only be expected if their longitudinal coherence 
can be substantially improved in the future.

A similar combination of lasers with massive linear arrangements of recycled 
HERA dipole magnets, but without a separating wall in the middle, 
might be used for the search for polarization effects induced by real and 
virtual axion production~\cite{Maiani:1986md}\footnote{For a variant of this idea, 
see Ref.~\cite{Cooper:1995zf}}. Indeed, 
axion production affects the polarization of laser light propagating in an external
magnetic field in two ways (see also Ref.~\cite{Raffelt:1987im}): 
the polarization vector of initially linearly polarized light will be
rotated by an angle $\epsilon_A$ and emerge with an elliptical polarization $\psi_A$,
\begin{eqnarray}
\label{ax_rot}
\epsilon_A &=& N_r\,\frac{g_{A\gamma}^2\,B^2\,\omega^2}{m_A^4}\,
\sin^2\left( \frac{m_A^2\,\ell}{4\,\omega}\right)\,
\sin 2\,\theta
\approx \frac{N_r}{16} \left( g_{A\gamma}\,B\,\ell \right)^2\,\sin 2\,\theta 
\,,
\\[1.5ex]
\label{ax_ell}
\psi_A &=& \frac{N_r}{2}\,\frac{g_{A\gamma}^2\,B^2\,\omega^2}{m_A^4}\,
\left[ \frac{m_A^2\,\ell}{2\,\omega} - \sin \left( \frac{m_A^2\,\ell}{2\,\omega}\right)
\right]\,
\sin 2\,\theta
\approx 
\frac{N_r}{96}\,\left( g_{A\gamma}\,B\,\ell \right)^2\,
\frac{m_A^2\,\ell}{\omega}\,
\sin 2\,\theta
\,,
\end{eqnarray}   
where $\theta$ is the angle between the light polarization direction and the 
magnetic field component normal to the light propagation vector. Thus, by measuring
both $\epsilon_A$ and $\psi_A$, one can determine the mass and the coupling of
the pseudoscalar~\cite{Maiani:1986md}. The ellipticity due to light-by-light
scattering in QED is much stronger than the one from axions, but calculable, 
\begin{equation}
\psi_{\rm QED} = 
\frac{N_r}{15}\,\alpha^2\,
\frac{B^2\,\ell\,\omega}{m_e^4}
\,\sin 2\,\theta
\,,
\end{equation}
where $m_e$ is the electron mass. 

A pioneering polarization experiment along these lines has been performed with the 
same laser and magnets described above exploited for the pilot photon regeneration 
experiment~\cite{Semertzidis:1990qc,Cameron:mr}. For $\ell = 8.8$~m, $B=2$~T, and $N_r=254$, 
an upper limit on the rotation angle $\epsilon_A< 3.5\times 10^{-10}$~rad 
was set, leading to a 
limit $g_{A\gamma}< 3.6\times 10^{-7}$~GeV$^{-1}$ at the 95\,\% confidence level, 
provided $m_A<8\times 10^{-4}$~eV. Similar limits have been set from the absence of 
ellipticity in the transmitted beam\footnote{The overall envelope of the current constraints from 
laser experiments~\cite{Semertzidis:1990qc,Ruoso:1992nx,Cameron:mr}, including both 
photon regeneration and polarization experiments, is shown
in Fig.~\ref{fig:ax_ph} and labelled by ``Laser''.}. 

The currently running PVLAS experiment~\cite{Brandi:2000ty}, consisting of a 
Fabry-P\'erot cavity of very high finesse ($N_r\gwig 90\, 000$) immersed in an 
intense magnetic dipole with $\ell =6.4$~m and $B=6.5$~T, 
will improve these bounds\footnote{A proposal similar to the PVLAS experiment is the 
Fermilab E-877 experiment~\cite{Nezrick:tx}.} on $g_{A\gamma}$ by a factor of $\sim 44$, 
as can be easily seen from Eqs.~(\ref{ax_rot}) and (\ref{ax_ell}). 
Similarly, a polarization experiment in the HERA tunnel, using $\approx 34$ 
of the HERA dipole magnets, is expected to improve the current limit on $\epsilon_A$ 
by a factor of 
\begin{equation}
(500/254)^{1/2}\,(5\,{\rm T}/2\,{\rm T})\,(340\,{\rm m}/8.8\,{\rm m})\approx 136\,,
\end{equation}
to $g_{A\gamma}\,\lwig\, 3\times 10^{-9}$~GeV$^{-1}$, for small $m_A$, Eq.~(\ref{eq:ax_coh}).  
Therefore, it seems that only with considerable efforts, in particular with respect to 
the improvement of the finesse of the cavity, a polarization experiment in the HERA tunnel 
can reach a sensitivity to the axion-photon coupling which is comparable to the 
one obtainable in a photon regeneration experiment of similar size in $B$ and $\ell$. 
Even a state-of-the-art ($N_r=500$) second stage polarization experiment in the 
TESLA XFEL tunnel can reach only a sensitivity of 
$g_{A\gamma}\,\lwig\, 5\times 10^{-10}$~GeV$^{-1}$ for small axion masses.  

In conclusion, we have shown that in the not so far distant future, when HERA has been 
decommissioned, there arises a unique opportunity for searches for light neutral (pseudo-)scalars 
by recycling a number of HERA's superconducting dipole magnets and using them for laser experiments 
exploiting photon regeneration or polarization. These experiments may reach a sensitivity in the 
coupling of the (pseudo-)scalar to two photons which is 
unprecedented in pure laboratory experiments and exceeds, for small masses, astrophysical and 
cosmological limits.  

\section*{Acknowledgements}
I wish to thank A. Melissinos, J\"org Ro\ss bach,  P. Schm\"user, Th. Tschentscher, 
and F. Willeke for valuable information concerning experimental and technical parameters. 
Encouraging discussions with W. Buchm\"uller, T. Greenshaw, F. Schrempp,   
D. Wyler, and P. Zerwas are also acknowledged.

\end{document}